\documentclass{llncs}

\usepackage{makeidx}  % allows for indexgeneration
\usepackage{microtype} 
\usepackage{siunitx} 
\usepackage{booktabs} 
\usepackage{cleveref}
\usepackage{amssymb} 
\usepackage{algorithm} 
\usepackage{algpseudocode} 
\usepackage{bbold}
\usepackage{enumerate}
\usepackage{cite}

\newcommand\eqref[1]{(\ref{#1})}

\begin{document}
\title{Optimal Design of Robust Combinatorial Mechanisms for Substitutable Goods}
\titlerunning{Optimal Design of Robust Mechanisms}  % abbreviated title (for running head)
%                                     also used for the TOC unless
%                                     \toctitle is used
%
\author{Maciej Drwal}
\authorrunning{Maciej Drwal} % abbreviated author list (for running head)
%
%%%% list of authors for the TOC (use if author list has to be modified)
\tocauthor{Maciej Drwal}
\institute{Department of Computer Science, Wroc\l{}aw University of Science and Technology,\\ Wybrze\.ze Wyspia\'nskiego 27, 50-370 Wroc\l{}aw, Poland\\
\email{maciej.drwal@pwr.edu.pl}}

\maketitle              % typeset the title of the contribution

\begin{abstract}
	In this paper we consider multidimensional mechanism design problem for selling discrete substitutable items to a group of buyers. Previous work on this problem mostly focus on stochastic description of valuations used by the seller. However, in certain applications, no prior information regarding buyers' preferences is known. To address this issue, we consider uncertain valuations and formulate the problem in a robust optimization framework: the objective is to minimize the maximum regret. For a special case of revenue-maximizing pricing problem we present a solution method based on mixed-integer linear programming formulation.
\end{abstract}


\section{Introduction}

We consider the following setup with a monopolist seller who wants to sell a set of substitutable items to a group of buyers. Each buyer has their own preferences over the offered items, and has a fixed demand. Each item is unique, and thus can be sold to one buyer (this is without the loss of generality, since the seller may be offering for sale several copies of the same item). 
%[??? what if copies are assigned different prices?]
The seller wants to determine the prices of items in order to maximize his revenue. However, his knowledge of buyers' preferences is limited. Indeed, the item valuations are private information of each buyer, who may want to strategically misreport them in order to receive items ahead of the others, or simply pay less than certain item is worth (according to the buyers' knowledge). In problems of this kind it is usually assumed that the seller has some probabilistic model of buyers' preferences, and wants to maximize his expected revenue, subject to the information he can extract from the available data. However, in many practical situations it is not possible to have any reliable statistical data to build such a model. Instead, a risk-averse seller may want to assume as little as possible about buyers' preferences, but enough to obtain a profit from the sale. One reasonable approach, motivated by a firm axiomatic basis \cite{milnor1951games, rosenhead1972robustness}, is to use robust optimization approach and assign prices to the items, so that the maximum regret of the revenue is minimized. We examine this approach in this paper for the considered combinatorial mechanism design problem.

%The considered problem is a variant of monopolist pricing for a set of discrete substitutable goods.
\subsection{Related Work}

The contribution of the paper would be summarized by contrasting it with the following related well-established research areas.

{\bf Combinatorial Auctions.} In combinatorial auctions buyers are described by valuation functions $v(S)$, where $S$ is a subset of items that seller has in the offer \cite{nisan1999algorithmic}. Typically, the goal of the auction is to allocate subsets of items to the buyers so that the sum of resulting utilities (the total welfare) is maximized. The computed allocation is a function of bids given by all buyers. A good design incentivizes truthful bidding. Combinatorial auction model is similar to the one presented in this paper, however, we consider not only the allocation problem, but also the problem of determining prices that maximize seller's revenue. Moreover, while combinatorial auction model allows for very general model of utility based on subsets of items, in this paper we consider a specific class of utility functions based on multidimensional valuation vectors.
%This can be thought of as if each buyer would know what is their utility of each bundle. 

{\bf Multidimensional Mechanism Design}. The problem considered in this paper is motivated by the works \cite{rochet1998ironing, manelli2007multidimensional, balcan2008item, cai2011optimal} and others, that build on the ideas initiated in \cite{myerson1981optimal}. In \cite{ulku2013optimal} a similar setup to ours is also presented, but with buyers' types being single-dimensional, and utility functions defined on subsets of items. %In contrast, we consider for each buyer independently valuated items, which requires multidimensional type space. 

However, the Bayesian approach proposed in these works is not adequate for the setup of interest in this paper. In particular, the Bayesian approach requires prior beliefs on buyers' valuations and/or historical data for parameter estimation. These assumptions do not always hold in practice, especially when a set of new products is introduced to the market, or the seller offers goods that are unique for the given buyers. Consequently, the {\it robust mechanism design} has already been proposed in recent publications \cite{bergemann2008pricing, handel2015robust, caldentey2015intertemporal, drwal2016}. However, these prior works focus on single item pricing, and to the best of our knowledge, min-max regret approach has not yet been applied to the multidimensional mechanism design for discrete goods. This paper investigates an approach that is complimentary to the Bayesian analysis, when robustness of solutions is needed.

\section{Problem Formulation}

In this section we give definitions of optimal combinatorial mechanism design and pricing problems for discrete substitutable items.

\subsection{Preliminaries}

We consider a seller which has $M$ substitutable items for sale, and $N$ concurrent buyers, each with demand for $D_i \leq M$ items. Each buyer is described by a vector of type ${\bf x}_i \in \Omega \subset \mathbb{R}^M$, where $x_{ij}$ is $i$th buyer's valuation of $j$th item. 

A ({\it direct-revelation}) {\it mechanism} is a pair of functions $(\phi, \psi)$, where $\psi$ maps a profile of buyers' valuations ${\bf X} = ({\bf x}_1, \ldots, {\bf x}_N)$ into a matrix of allocations of items ${\bf Q} = ({\bf q}_1, \ldots, {\bf q}_N)$, and $\phi$ maps allocations of items ${\bf Q}$ into the vector of payments ${\bf p}$ that each $i$th buyer makes for receiving items according to the allocation ${\bf Q}$. Function $\phi$ is called the {\it payment} function, and $\psi$ is called the {\it allocation} function.

In this paper we only consider mechanisms of a specific kind. The allowed allocation matrices ${\bf Q}$ are only such that each buyer receives up to $D_i$ items, and each item is assigned to up to a single buyer. Formally, $\psi: \Omega^N \rightarrow \mathcal{Q}$, where:
$$
	\mathcal{Q} = \{ {\bf Q} \in \{ 0, 1 \}^{MN} : \;\; \sum_{i=1}^N q_{ij} \leq 1, \;\; \sum_{j=1}^M q_{ij} \leq D_i \}.
$$
Moreover, prices of items are independent of the buyer's index, thus the payment function $\phi : \mathcal{Q} \rightarrow \mathbb{R}_+^N$ can be defined by a vector ${\bf p} = (p_1, \ldots, p_M)$, so that $\phi({\bf Q}) = (\phi_1({\bf Q}), \ldots, \phi_N({\bf Q}))$, where:
$$
	\phi_i({\bf Q}) = \sum_{j=1}^N p_j q_{ij}.
$$
The seller's revenue is defined as the sum of payments all the buyers make, given the type profile ${\bf X}$:
$$
	r({\bf p}, \psi; {\bf X}) = \sum_{i=1}^N \sum_{j=1}^M p_j \psi_{ij}({\bf X}).
$$
Buyer's utility is defined by a quasi-linear function: $u_i({\bf p}, {\bf q}_i, {\bf x}_i) = {\bf q}_i \cdot \left( {\bf x}_i - {\bf p} \right)$. A mechanism is called {\it incentive-compatible} (IC), if the allocations corresponding to the true valuations maximize utilities of each buyer. Then a dominant strategy of each buyer is to report his true (private) valuations to the seller. Formally, mechanism $({\bf p}, \psi)$ is IC if:
$$
	\forall_i \; \forall_{{\bf x}' \in \Omega} \;\;\; u_i({\bf p}, \psi_i({\bf X}), {\bf x}_i) \geq u_i({\bf p}, \psi_i({\bf x}', {\bf X}_{-i}), {\bf x}_i),
$$
which in the considered case is equivalent to:
\begin{equation}\label{IC-con}
	\forall_i \; \forall_{{\bf x}' \in \Omega} \;\;\; \left( \psi_i({\bf X}) - \psi_i({\bf x}', {\bf X}_{-i}) \right) \cdot \left( {\bf x}_i - {\bf p} \right) \geq 0.
\end{equation}
Notation $({\bf x}', {\bf X}_{-i})$ denotes the matrix ${\bf X}$ with $i$-th column ${\bf x}_i$ replaced with ${\bf x}'$.

It is further assumed that buyers are {\it individually rational} (IR), which means that they participate in the mechanism only when the utility gained is at least equal to a certain threshold $u_{i,0}$:
\begin{equation}\label{IR-con}
	\forall_i \; u_i({\bf p}, \psi_i({\bf X}), {\bf x}_i) \geq u_{i,0}.
\end{equation}
Henceforth, we normalize the threshold for all buyers to zero, i.e., $u_{i,0} = 0$ for $i=1,\ldots,N$. We denote the set of all {\it feasible} mechanisms by:
\begin{equation}\label{M-set}
	\mathcal{M} = \left\{ ({\bf p}, \psi): \; {\bf p} \in \mathbb{R}_+^M, \; \forall_{{\bf X} \in \Omega^N} \;\; \psi({\bf X}) \in Q  \textrm{ and satisfy } \eqref{IC-con} \textrm{ and } \eqref{IR-con} \right\}.
\end{equation}

\subsection{Robust Mechanism Design}

We assume that vectors of buyers' types are not known to the seller. Instead, there is a known set of possible type scenarios $\mathcal{S} \subset \Omega^N$. Two most widely considered special cases of $\mathcal{S}$ are discrete scenario sets and interval uncertainty sets. The {\it discrete scenario} set is defined as $ \mathcal{S} = \{ {\bf X}^{(1)}, \ldots, {\bf X}^{(S)} \}$. The {\it interval uncertainty} set is defined as $\mathcal{S} = \{ [x_{ij}^-, x_{ij}^+] \subset \mathbb{R}_+ : \; i=1,\ldots,N, \; j=1,\ldots,M \}$.

For a fixed scenario ${\bf X} \in \mathcal{S}$ the regret of a mechanism $({\bf p}, \psi)$ is defined as:
$$
	R({\bf p}, \psi; {\bf X}) = \left( \max_{({\bf p}', \psi') \in \mathcal{M}} r({\bf p}', \psi'; {\bf X}) \right) - r({\bf p}, \psi; {\bf X}),
$$
i.e., the difference between the revenue generated by an optimal mechanism for known valuations ${\bf X}$, and the revenue generated by the mechanism $({\bf p}, \psi)$. The objective of the seller is to determine a mechanism $({\bf p}^*, \psi^*)$ that minimizes the maximum regret over the set of scenarios:
\begin{equation}\label{rob-prob}
	({\bf p}^*, \psi^*) \in \arg \min_{({\bf p}, \psi) \in \mathcal{M}} \max_{{\bf X} \in \mathcal{S}} R({\bf p}, \psi; {\bf X}).
\end{equation}

The problem of finding an optimal robust mechanism \eqref{rob-prob} can be seen as three-level optimization problem, in which we minimize (over the space of all feasible mechanisms) the objective function involving unconstrained optimization over the set of all type profiles ${\bf X} \in \mathcal{S}$, with objective function that contains a term involving constrained optimization of revenue for a fixed type profile (again over the space of all feasible mechanisms). The innermost sub-problem of finding revenue-maximizing prices and allocations is usually referred to as the {\it deterministic} problem, as it assumes fixed ${\bf X}$ given as an input. The outermost problem will be referred to as the {\it robust} problem.

Similarly to the Bayesian variant of the multidimensional mechanism design, the robust version of the problem is difficult to solve. Since we model buyers' valuations using subsets of multidimensional real space, the set \eqref{M-set} of feasible mechanisms $({\bf p}, \psi)$ contains allocations $\psi$ expressed as functions on possibly continuous domain. This makes the problem \eqref{rob-prob} a functional optimization problem (this is also the case of general optimal mechanism design problem \cite{rochet1998ironing}). Note that, given the total function solution $\psi$, the seller would be able to optimally assign items for all buyers' types ${\bf X} \in \Omega^N$, and that would cover also non-truth telling buyers (which do not comply to the incentive-compatibility assumption).

\subsection{Robust Pricing}

Instead of computing an optimal robust mechanism, in some applications it may be enough for the seller to determine only an optimal robust set of prices. In this slightly simplified solution concept, buyers' reported types will not map directly into an allocation rule. Instead, reported types may be the basis for constructing the uncertainty set $\mathcal{S}$, given as an input to the seller. If buyers are aware of the fact that the seller will price the items so that each buyer can receive their utility-maximizing item, truth telling will remain to be an equilibrium strategy. Note that in this approach the solution is also more resistant to the buyers' types misreporting caused by their inherent uncertainty of item valuations (i.e., {\it non-strategic} misreporting).

It is enough to consider unit demands $D_i = 1$ for all buyers $i=1,\ldots,N$, since if in the original problem a buyer have arbitrary demand $D_i > 1$, we substitute them with $D_i$ identical buyers of unit demands.
Consequently, we can assume that matrix ${\bf X}$ is square. Thus if there are less items than buyers we add zero-value dummy items, and interpret matching a dummy item to buyer as not allocating any items to that buyer. If there are less buyers than items we add all-zero valuations buyers, and interpret assigning to such a buyer as not selling an item. Let $K = \max\{ M, N \}$.

Consequently, the optimal (deterministic) pricing problem can be defined for any ${\bf X} \in \Omega^{K}$ as the follows:
\begin{equation}\label{pricing-prob}
	\max_{{\bf p} \geq 0, {\bf Q} \in \mathcal{Q}} r({\bf p}, {\bf Q}),
\end{equation}
subject to:
\begin{equation}\label{IC-con-1}
	\forall_i \; \forall_j \;\;\; \sum_{k=1}^K q_{jk} (x_{jk} - p_k) \geq \sum_{k=1}^K q_{ik} (x_{jk} - p_k),
\end{equation}
\begin{equation}\label{rob-con-4}
	\forall_j \;\;\; \sum_{k=1}^K q_{jk} (x_{jk} - p_k) \geq 0,
\end{equation}
where $r({\bf p}, {\bf Q}) = \sum_{i=1}^N {\bf p} \cdot {\bf q}_i$ is the revenue.

Constraint \eqref{IC-con-1} forces the assignment of items to be utility-maximizing for each buyer. The problem \eqref{pricing-prob}--\eqref{rob-con-4} is a mixed-integer nonlinear program. Introducing new variables $u_i$, and substituting them for each $i$th buyer's utility $u_i = \sum_{k=1}^K q_{ik}(x_{ik} - p_k)$, we can transform the nonlinear formulation into a mixed-integer linear one as follows:
\begin{equation}\label{pricing-prob-1}
	\max_{{\bf u} \geq 0, {\bf Q} \in \mathcal{F}({\bf u}, {\bf X})} \sum_{i=1}^K \left(  \sum_{j=1}^K q_{ij}x_{ij} - u_i \right),
\end{equation}
where:
\begin{equation}\label{F-set}
\mathcal{F}({\bf u}, {\bf X}) = \{ {\bf Q} \in \mathcal{Q} : \; \forall_i, \forall_j \;\; u_j - u_i \geq \sum_{k=1}^K q_{ik}( x_{jk} - x_{ik})  \}.
\end{equation}
The equivalence of the constraint sets given by \eqref{IC-con-1}--\eqref{rob-con-4} and $\mathcal{F}$ can be seen by observing that the lefthand side of \eqref{IC-con-1} is equal to $u_j$, thus by subtracting $u_i$ from both sides of this constraint, and expanding the right-hand side, we obtain the inequality defining \eqref{F-set}. We note that this new constraint is a special case of more general condition that incentive-compatible allocations must correspond to utility functions that are convex continuous (see, e.g., \cite{manelli2007multidimensional}). These inequalities are a discretized equivalent of the convexity constraint imposed on utility functions. Constraint \eqref{rob-con-4} becomes a standard non-negativity constraint ${\bf u} \geq 0$. Given optimal solution $({\bf u}^*, {\bf Q}^*)$, the vector of optimal prices ${\bf p}^*$ can be computed from the definition of utility.

Before we give the robust formulation of this problem, we need to observe that the feasibility of a particular solution $({\bf u}, {\bf Q})$ depends on the actual scenario ${\bf X}$. Most notions of robustness require that solution should be given before the true scenario is realized \cite{ben2009robust}. Consequently, we will require from the robust solution that it is unconditionally feasible, that is, regardless of the scenario. Formally, solution must be {\it robust feasible}, i.e., from the set:
$$
\mathcal{R} = \bigcap_{X \in \mathcal{S}} \{ ({\bf u}, {\bf Q}) : \; {\bf u} \geq 0, \; {\bf Q} \in \mathcal{F}({\bf u}, {\bf X}) \}.
$$
Note that in many cases robust feasible solutions require that not every item is sold, i.e., matrix ${\bf Q}$ does not have a full rank.
The robust formulation of the pricing problem is the following:
\begin{equation}\label{rob-pricing}
	\min_{({\bf u}, {\bf Q}) \in \mathcal{R}} \max_{{\bf X} \in \mathcal{S}} \left( \max_{({\bf u}', {\bf Q}') \in \mathcal{F}({\bf u}', {\bf X})} \sum_{i=1}^K \left( \sum_{j=1}^K q'_{ij} x_{ij} - u'_i \right) - \sum_{i=1}^K \left(  \sum_{j=1}^K q_{ij} x_{ij} + u_i \right) \right).
\end{equation}

\section{Solution Method for Interval Uncertainty}

In this section we present an algorithm for solving \eqref{rob-pricing} for the interval uncertainty case of valuations set $\mathcal{S}  = \{ [x_{ij}^-, x_{ij}^+] : \; i, j \in \{ 1,\ldots,K \} \}$. The algorithm is based on Benders cut generation method. It makes use of the mixed-integer linear program for deterministic pricing problem \eqref{pricing-prob-1}--\eqref{F-set}. A similar solution method was used for the robust assignment problem in \cite{pereira2011exact}, however problem \eqref{rob-pricing} introduces additional constraints on feasible assignments.

The algorithm proceeds as follows:
\begin{enumerate}
	\item Let $\mathcal{A} \leftarrow \emptyset$, $LB \leftarrow -\infty$, $UB \leftarrow +\infty$.
	\item Solve:
	$$
		\min_{{\bf Q} \in \mathcal{Q}, {\bf u} \geq 0, \theta \geq 0} \left( \theta - \sum_{i=1}^K \left( \sum_{j=1}^K q_{ij} x_{ij}^- - u_i \right) \right),
	$$
	subject to:
	$$
	\forall_{i,j} \;\; u_j - u_i \geq \sum_{k=1}^K q_{ik} (x_{jk}^+ - x_{ik}^-)
	$$
	and:
	$$
	\forall_{({\bf u}', {\bf Q}') \in \mathcal{A}} \;\; \theta \geq \sum_{i=1}^K \left( \sum_{j=1}^K q_{ij}' \left( x_{ij}^+ + (x_{ij}^- - x_{ij}^+) q_{ij} \right) - u_i' \right).
	$$
	Let $(\hat{\bf u}, \hat{\bf Q}, \hat{\theta})$ be an optimal solution, and $\hat{v}$ be the value of this solution. \label{main-loop}
	\item If $\hat{v} > LB$ then $LB \leftarrow \hat{v}$.
	\item Solve the deterministic problem \eqref{pricing-prob-1}--\eqref{F-set} for scenario:
	$$
		x_{ij} = \left\{ \begin{array}{cc} x_{ij}^-, & \hat{q}_{ij} = 1, \\ x_{ij}^+, & \hat{q}_{ij}=0. \\ \end{array} \right.
	$$
	Let $(\bar{u}, \bar{\bf Q})$ be an optimal solution.
	\item Compute value $\tilde{v}$ of solution $(\hat{\bf u}, \hat{\bf Q}, \bar{\bf u}, \bar{\bf Q})$ for \eqref{rob-pricing}. If $\tilde{v} < UB$ then $UB \leftarrow \tilde{v}$.
	\item If $UB \leq LB$ then STOP.
	\item Add $(\bar{\bf u}, \bar{\bf Q})$ to the set $\mathcal{A}$ and go to Step \ref{main-loop}.
\end{enumerate}

The procedure starts from relaxing constraints that define set $\mathcal{F}({\bf u}, {\bf Q})$, restricting the solution to be robust feasible, $({\bf u}, {\bf Q}) \in \mathcal{R}$. An initial solution is found by optimizing the relaxed problem. Such a solution, however, is usually not feasible for the problem \eqref{rob-pricing}, but only provides a lower bound (variable $LB$). However, given this solution, we determine the worst-case scenario ${\bf X} \in \mathcal{S}$, which is an extreme scenario, consisting of only lower or upper interval bounds for each valuation. A deterministic pricing problem \eqref{pricing-prob-1}--\eqref{F-set} is then solved for that worst-case scenario, which corresponds to the inner maximization sub-problem in \eqref{rob-pricing}. However, since it is computed only for one particular choice of optimization variables $({\bf u}, {\bf Q})$, the solution of this sub-problem gives only a lower bound on the first term in the objective function of \eqref{rob-pricing}, which is represented by a new optimization variable $\theta$. Given a solution of the sub-problem, we obtain a feasible solution of \eqref{rob-pricing}, which provides an upper bound on an optimal one (variable $UB$). We create a Benders cut from the solution of the sub-problem, and add its indexing variable to the set $\mathcal{A}$. Note that in order to completely describe the feasible set of \eqref{rob-pricing} we would require all robust feasible solutions to be contained in $\mathcal{A}$. But then $\mathcal{A}$ would have exponential size. However, it is very often that most of these constraints are superfluous, and an optimal solution of \eqref{rob-pricing} can be found by taking only very small fraction of these constraints into consideration.

\subsection{Experimental Results}

For an experimental evaluation we have prepared interval uncertainty sets $\mathcal{S}$ by randomly generating lower bounds of valuations uniformly from range $[x_{\min}^-, x_{\max}^-]$, and then for the corresponding upper bound by adding number uniformly from range $[0, \Delta]$, independently for each buyer and item pair. Table \ref{tab:0} shows the summary of results. Problem instances were solved using cut-generating algorithm described in the previous section. For each problem size $K$ the values given are averaged over 10 repetitions of experiment. Columns contain, respectively: the problem size (number of buyers and items), value of an optimal regret, revenue generated by robust optimal solution, resulting welfare (sum of buyers' utilities), number of items sold in the robust optimal solution, and approximate computation time in seconds. Solutions for $K \leq 30$ are optimal with absolute error allowed $0.05$. Solutions for $K \geq 40$ were computed with time limit of 1 hour, and in most cases no optimal solution was found until then, thus results for best feasible solutions are reported. In a vast majority of cases it was enough to generate only 2--4 cuts in order to find a robust optimal solution. 

Notice that for a large number of items and competing buyers the revenue generated by robust solutions starts to decrease, after reaching its highest level. We may conclude that, depending on the range of uncertainty sets, there is a certain number of items that the seller should try to sell concurrently, in order to safely profit from the competition between buyers. However, above that threshold the buyers' utilities quickly decrease and so the guaranteed revenue drops down (indicated by a robust solution), thus it is best to reduce the size of the product line.

\begin{table}[ht!]
	\centering
	\caption{Experimental results. Highest values of average revenue generated by robust solutions are marked in bold.}\label{tab:0}
\begin{tabular}{|c|c|c|c|c|c|}
	\hline
	 % $n$ & \multicolumn{2}{c|}{``flat'' prices} & \multicolumn{2}{c|}{basic mechanism} & \multicolumn{2}{c|}{2-stage mechanism}\\
	 %     & cancelled & income & cancelled & income & cancelled & income \\
	 $\;\;\; K \;\;\;$ & optimal regret & robust revenue & robust welfare & sold items & time \\
	\hline

	 				   & \multicolumn{5}{c|}{$x_{\min}^-=10$, $x_{\max}^-=500$, $\Delta = 30$} \\
	\hline
	 5 & 111.27 & 1452.45 & 133.72 & 4.18 & 11.94 \\
	 10 & 623.27 & 3518.00 & 96.00 & 7.91 & 17.79 \\
	 20 & 3116.27 & 5687.00 & 72.00 & 12.00 & 287.04 \\
	 30 & 7464.82 & {\bf 6393.45} & 57.64 & 13.27 & 7565.10 \\
	 40 & 13324.82 & 5505.18 & 19.09 & 11.27 & 10946.47 \\
	 50 & 18785.75 & 5282.00 & 11.00 & 10.75 & 10262.33 \\
	%500 & 353 & 1497.1 & 322 & 2774.81 & 311 & 4809.03 \\ 
	 %1000 & 640 & 4582.0 & 755 & 4667.98 & 766 & 4044.01\\
	\hline
		               & \multicolumn{5}{c|}{$x_{\min}^-=10$, $x_{\max}^-=500$, $\Delta = 50$} \\
	\hline
	  5 & 200.09 & 1390.00 & 84.18 & 3.73 & 11.73 \\ 
	 10 & 1148.73 & 3037.36 & 116.18 & 6.91 & 16.60 \\
	 20 & 5024.36 & {\bf 4012.45} & 50.18 & 8.45 & 203.89 \\
	 30 & 10534.55 & 3660.55 & 15.55 & 7.55 & 8451.96 \\
	 40 & 16941.36 & 2349.09 & 9.54 & 4.81 & 8452.51 \\
	 50 & 22709.09 & 1876.00 & 0.45 & 3.82 & 10433.85 \\
	\hline
		               & \multicolumn{5}{c|}{$x_{\min}^-=100$, $x_{\max}^-=500$, $\Delta = 50$} \\
	\hline
	 5 & 284.27 & 1453.27 & 90.55 & 3.73 & 13.18 \\
	10 & 1611.36 & 2736.64 & 80.36 & 6.09 & 14.11 \\
	20 & 6126.09 & {\bf 3161.00} & 28.00 & 6.55 & 145.86 \\
	30 & 12098.45 & 2365.91 & 2.91 & 4.82 & 6733.55 \\
	40 & 17864.73 & 1695.55 & 2.91 & 3.45 & 9937.67 \\
	50 & 23881.71 & 987.43 & 0.00 & 2.00 & 10466.73 \\
	\hline
		               & \multicolumn{5}{c|}{$x_{\min}^-=100$, $x_{\max}^-=1000$, $\Delta = 50$} \\
	\hline
	5 & 175.82 & 2937.45 & 269.91 & 4.27 & 10.94 \\
	10 & 742.64 & 7422.36 & 329.09 & 8.55 & 15.97 \\
	20 & 5437.09 & 12043.00 & 218.09 & 12.82 & 361.68 \\
	30 & 12198.00 & {\bf 15526.60} & 159.10 & 16.10 & 9316.22 \\
	40 & 24506.00 & 12933.67 & 54.33 & 13.33 & 9345.68 \\
	50 & 31608.00 & 6151.00 & 86.00 & 16.50 & 10804.96 \\
	\hline
\end{tabular}
\end{table}

\subsection{Algorithm for Deterministic Sub-Problem}\label{sec:alg-deter}

Finally, we present a fast heuristic for solving sub-problem \eqref{pricing-prob-1}--\eqref{F-set} for a fixed type profile. It can be used as a sub-routine in the outer problem of minimizing the maximum regret, when the problem size is prohibitive for applying exact algorithm from the previous subsection. Note that this problem can be also solved directly using mixed-integer linear programming. The presented procedure is based on solving assignment problem, using e.g., Hungarian algorithm \cite{papadimitriou1982combinatorial}.

Observe that the revenue-maximizing item allocation is upper-bounded by the value of maximum weight matching of items to buyers. Indeed, if ${\bf Q} \in \mathcal{Q}$ is one-to-one matching, and $p_j$ is the price of $j$th item for some (unique) buyer $i$, such that $Q_{ij} = 1$, then the revenue ${\bf Q} \cdot {\bf p}$ is also the value of the matching. However, the vector of prices ${\bf p}$ corresponding to the maximum weight matching may violate constraints defining \eqref{F-set}. The idea of solution algorithm is to adjust the prices so that to make sure that these constraints are satisfied, while at the same time keeping the assumed allocation ${\bf Q}$ corresponding to the maximum weight matching. Note however, that such allocation may not be optimal for any choice of prices ${\bf p}$.

% Observe that the revenue-maximizing item allocation is upper-bounded by the value of maximum weight matching of items to buyers. Indeed, if $\psi({\bf X}) \in Q$ is one-to-one matching, and $p_j$ is the price of $j$th item for some (unique) buyer $i$, such that $\psi_{ij}({\bf X}) = 1$, then the revenue $\psi({\bf X}) \cdot {\bf p}$ is also the value of the matching. However, the vector of prices ${\bf p}$ corresponding to the maximum weight matching may violate constraints defining \eqref{F-set}. The idea of solution algorithm is to adjust the prices so that to make sure that these constraints are satisfied, while at the same time keeping the allocation $\psi$ corresponding to the maximum weight matching.

The algorithm is the following:
\begin{enumerate}
	\item Find a maximum weight assignment for the matrix ${\bf X}$ of buyers' valuation profile (each column ${\bf x}_i$ corresponds to buyer's $i$ valuations, and each row ${\bf X}^T_j$ corresponds to an item $j$). Denote the matching by permutation matrix ${\bf A}$.
	\item Let ${\bf p} \leftarrow (p_1, \ldots, p_M)$, where $p_j = x_{ij}$, such that $A_{ij}=1$ for some $i=1,\ldots,N$.
	\item For each buyer $n=1,\ldots,N$:\label{step-3}
	\begin{enumerate}
		\item Let $U_i$ be the utility of matched item $j_0$, i.e., $U = x_{ij_0} - p_{j_0}$, where $A_{ij_0}=1$.
		\item Let $\hat{U}_i$ be the maximum utility of buyer $i$ given the current prices, i.e., $\hat{U}_i = x_{ij'} - p_{j'}$, where $j'$ maximizes $\{ x_{ij} - p_j : \; j = 1,\ldots,M \}$. Here $j'$ is the item that client prefers under the current prices.
		\item If $\hat{U}_i > U_i$ then $p_{j_0} \leftarrow p_{j_0} - \hat{U}_i - U_i$.
		\item If $p_{j_0} < 0$ or $U_{j_0} < 0$ set $A_{ij_0} = 0$ (i.e., item $j_0$ is not sold).\label{step-recomp}
	\end{enumerate}
	\item Check if $\hat{U}_i = U_i$ for all buyers $i=1,\ldots,N$. If not, then go to Step \ref{step-3}.
\end{enumerate}

The procedure can be further improved. For example, assignment matrix $A$ can be recomputed when initial item match turns out to be infeasible for a given buyer in Step \ref{step-recomp} (in a new valuation matrix we would set the buyer's valuation of that item to zero).

\section{Conclusions and Further Work}

The problem of optimal mechanism design, while relatively easy for single-dimensional types, becomes more difficult to solve in the case of multidimensional type space and specific constraints regarding allowed allocations of goods. Different formulations using Bayesian models of valuation uncertainty are subject to active ongoing research. In this paper we presented formulation in which no prior beliefs regarding multidimensional valuations are assumed, and the objective is to design a robust mechanism in a min-max regret sense. We provided a cut-generation based algorithm for solving a special case of pricing mechanism for interval uncertainty. 

There are many directions of interesting further research. One is to assume budget-constrained buyers with arbitrary demands. Then we would obtain a formulation of the deterministic problem that contains a generalized assignment problem as a special case, thus the problem becomes even more complex. Another idea is to introduce a regulated monopoly: in such case it is required to design a mechanism that maximizes a weighted sum of seller's revenue and buyers' welfare. Finally, more efficient solution algorithms would allow to extend the range of possible practical applications for the class of mechanisms under consideration.

%(Idea: if buyers were budget-constrained then we would get generalize assignment problem with demands equal to budget, resource requirement equal to price and equal to the profit in objective function. However then even the deterministic problem becomes NP-hard.)

%(Idea: regulated monopoly: maximize weighted sum of revenue and welfare)

% \bibliographystyle{elsarticle-num}
% \bibliography{main}

\end{document}